\documentclass{jfm}
\usepackage{graphicx}
\usepackage{epstopdf, epsfig}
\usepackage{xcolor}
\usepackage{amsmath}

\shorttitle{Viscous resuspension of non-Brownian suspensions}
\shortauthor{E. d'Ambrosio, F. Blanc, E. Lemaire}

\title{Viscous resuspension of non-Brownian particles: determination of the concentration profiles and particle normal stresses.}

\author{Enzo d'Ambrosio\aff{1},
Fr\'ed\'eric Blanc\aff{1}
\and Elisabeth Lemaire\aff{1}\corresp{\email{elisabeth.lemaire@unice.fr}}}

\affiliation{\aff{1}Universit\'e C\^{o}te d'Azur, CNRS, InPhyNi-UMR 7010, 06108 Nice Cedex 2, France}

\begin{document}

\maketitle

\begin{abstract}
We perform local measurements of both the shear rate and the particle fraction to study viscous resuspension in non-Brownian suspensions. A suspension of PMMA spherical particles dispersed in a lighter Newtonian fluid (\textit{Triton X100}) is sheared in a vertical Couette cell. The vertical profiles of the particle volume fraction are measured for Shields numbers ranging from $10^{-3}$ to $1$, and the variation in the particle normal stress in the vorticity direction of the particle fraction is deduced.
\end{abstract}


\section{Introduction}
Understanding the flow properties of concentrated suspensions is a real challenge in the development of many industrial products (e.g., solid propellant rocket motors and fresh concrete) and in the description of various environmental flows (e.g., torrential lava, mud flows, and submarine slides). Among other transport properties, shear-induced particle migration has received increasing attention in recent decades. Particle migration can be due to inertial effects \citep{segre1962behaviour} but also occurs at low Reynolds numbers, for instance, in a Poiseuille flow in which the particles tend to migrate towards the centre of the channel \citep{koh1994experimental,hampton1997migration,butler1999imaging,snook2016dynamics}, in wide-gap Couette flow towards the outer cylinder \citep{graham1991note,abbott1991experimental,chow1994shear,sarabian2019fully} and outward in cone-and-plate geometry \citep{chow1995particle}. Another typical example of shear-induced migration is viscous resuspension whereby an initially settled layer of negatively buoyant particles expands vertically when a shear flow is applied. Viscous resuspension has been observed for the first time by \cite{gadala1979rheology} and later explained by  \cite{leighton1986viscous} and \cite{acrivos1993shear}, who demonstrated that the height of the resuspended particle layer results from the balance between a downward gravitational flux and an upward shear-induced diffusion flux. The authors studied the resuspension of various particles (different sizes and densities) in two different liquids (different viscosities and densities) sheared in a cylindrical Couette device. They measured the height of the resuspended layer of particles, $h_s$, as a function of the shear rate and showed that the difference between $h_s$ and $h_0$ (i.e., the initial sediment height) normalized by $h_0$ was a function of only the Shields number defined as the ratio between viscous and buoyancy forces:
\begin{equation}
    \dfrac{h_s-h_0}{h_0}=f(A) \mbox{ with } A=\dfrac{9}{2}\dfrac{\eta_0 \dot{\gamma}}{\Delta \rho g h_0}
    \label{A}
\end{equation}
Their experimental results were found to be in very good agreement with the diffusive flux model developed by \cite{leighton1986viscous}.
Later, \cite{zarraga2000characterization} revisited the results of \cite{acrivos1993shear} to determine the particle normal stress in the vorticity direction, $\Sigma_{33}^p$, from the height of the resuspended layer of particles by writing the Cauchy momentum balance in the vertical direction:
\begin{equation}
    \dfrac{\partial  \Sigma_{33}^p}{\partial z}=\Delta \rho g \phi
    \label{Cauchy}
\end{equation}
Then, a relation between $\Sigma_{33}^p$ and the particle volume fraction at the bottom, $\phi_0$, is obtained by the integration of Eq. \ref{Cauchy} from the interface between the suspended layer and the clear liquid at the bottom together with the equation of particle number conservation. The relationship between particle normal stress and shear-induced migration (or resuspension) has been the subject of several studies and is still an active area of investigation \citep{nott1994pressure,mills1995rheology,morris1998pressure,morris1999curvilinear,deboeuf2009particle,lhuillier2009migration,nott2011suspension,ovarlez2013migration}.
The suspension balance model proposed by \cite{morris1999curvilinear} and refined by \cite{lhuillier2009migration} and \cite{nott2011suspension} offers a promising framework for modelling shear-induced particle migration, but it suffers from a relative lack of experimental data on particle normal stresses.
In addition to the above-cited work of \cite{zarraga2000characterization}, who used the viscous resuspension experiment of \cite{acrivos1993shear} to deduce $\Sigma_{33}^p$ for particle volume fractions ranging from $0.3$ to $0.5$, \cite{deboeuf2009particle} determined $\Sigma_{33}^p$ for particle volume fractions ranging from $0.3$ to $0.5$ through the measurement of the pore pressure in a cylindrical Couette flow. \cite{boyer2011unifying} used a pressure-imposed shear cell to measure $\Sigma_{22}^p$ in the range $\phi \in [0.4,0.585]$, and \cite{dbouk2013normal} determined $\Sigma_{22}^p$ in the range $\phi \in [0.3,0.47]$ through the measurement of both the total stress $\Sigma_{22}$ and the pore pressure. See \citep{guazzelli2018rheology} for a review. All of these studies show a linear relationship between the particle normal stress components and the shear rate, but recently, \cite{saint2019x} performed X-ray radiography experiments on viscous resuspension that revealed a non-linear relationship with the shear rate.

In this paper, we present the experimental results of viscous resuspension in a Couette device in which the local particle volume fraction and the local shear rate are measured by optical imaging. $\Sigma_{33}^p$ is obtained by integrating Eq.\ref{Cauchy} from the interface between the clear fluid and the resuspended layer to any height $z$ below the interface. These experiments present the dual advantage that $\Sigma_{33}^p$ can be determined for a wide range of particle fractions and that the local shear rate can be measured to accurately test the scaling of particle normal stresses with shear rate.

\section{Materials and Methods}
\subsection{Suspension and device}
PMMA spheres (Arkema BS572), $2a=268\pm 25\,\mu m$ in diameter and $1.19\, 10^{3}\pm 10\, kg/m^3$ in density, are used. The particles are dispersed in \textit{Triton X 100} to which a small amount of a fluorescent dye (\textit{Nile Blue A}, 
REF) 
is added. This mixture is Newtonian with a viscosity of $\eta_0=0.34
\pm0.02\, Pa.s$ and of density $1.06\,10^3\pm 10\, kg/m^3$ at $T=23^oC$. The characteristic settling velocity of the particles is then $V_S= 2/9\, \Delta \rho g/ \eta\approx 20\,\mu m/s$. The liquid and the particles are chosen to have roughly the same refractive index, $ 1.49$, and accurate index matching is achieved by tuning the temperature of the chamber that contains the rheometer.

\begin{figure}
    \centering
    \includegraphics[width=0.6\textwidth]{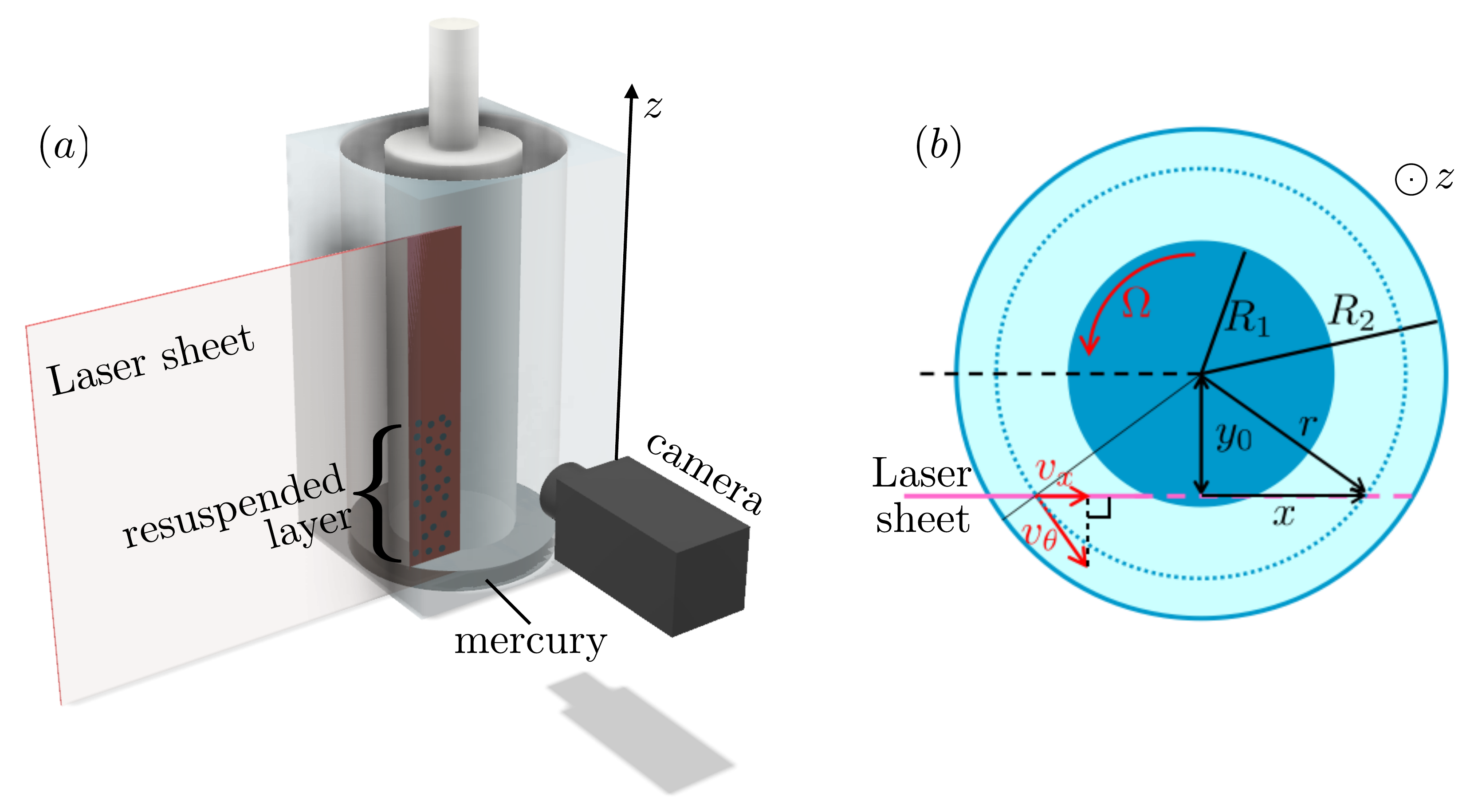}
\caption{a) Sketch of the experimental device. b) View from above. The vertical laser sheet is shifted by an offset of length $y_0$ from the radial position. $x$ is the horizontal position in the laser sheet, and $z$ is the vertical position. $z=0$ is set by the mercury/suspension interface.}
    \label{fig:schema}
\end{figure}

The resuspension experiments are conducted in a Couette cell made of PMMA mounted on a controlled-stress rheometer (Mars II, Thermofisher) (see figure \ref{fig:schema}(a)). The rotor has a radius $R_1 = 19\, mm$, and the stator has a radius $R_2 = 24\, mm$. Thus, the gap is much larger than the particle diameter ($(R_2-R_1)/a \approx 37$) but is small enough for the shear stress variation to be weak: $\Sigma_{12}(R_1)/\Sigma_{12}(R_2)=R_2^2/R_1^2\approx1.6$. Thus, radial shear-induced particle migration is expected to be weak.

The bottom of the Couette cell is filled with mercury to prevent the particles from migrating out of the gap (under the cup) and to  maximize slip at the suspension/bottom interface to have a shear rate as homogeneous as possible inside the gap. The suspension is poured into the rheometer cell and illuminated by a thin vertical laser sheet (thickness $\approx 50\, \mu m$) offset by $y_0=16.2\, mm$ from the radial plane (see figure \ref{fig:schema}(b)). A camera (IDS, nominal frequency 33 Hz, full resolution $4104 \times 2174 \, px^2$) is placed at $90^o$ to the enlightened plane. The accurate matching of the refractive index, the thinness of the laser sheet and the resolution of the camera allow the recording of high-quality images with a resolution of $30\,px$ per particle.

\subsubsection{Experimental procedure}
In this paper, we will focus on the steady state of resuspension obtained for various angular velocities of the rotor, $\Omega$: $0.3,\ 0.5,\ 1,\ 2,\ 5,\ 10,\ 20,\ 30,\ 40$ and $60$ rounds per minute ($rpm$). For all these angular velocity values, the Reynolds number ($\mathcal{R}e=\rho \Omega R_1(R_2-R_1)/\eta$) is less than $1$, and the P\'eclet number ($\mathcal{P}e=6\pi \eta a^3\dot{\gamma}/k_B T$) is very large ($\mathcal{P}e >10^{8} $).

\begin{figure}
    \centering
    \includegraphics[width=1\textwidth]{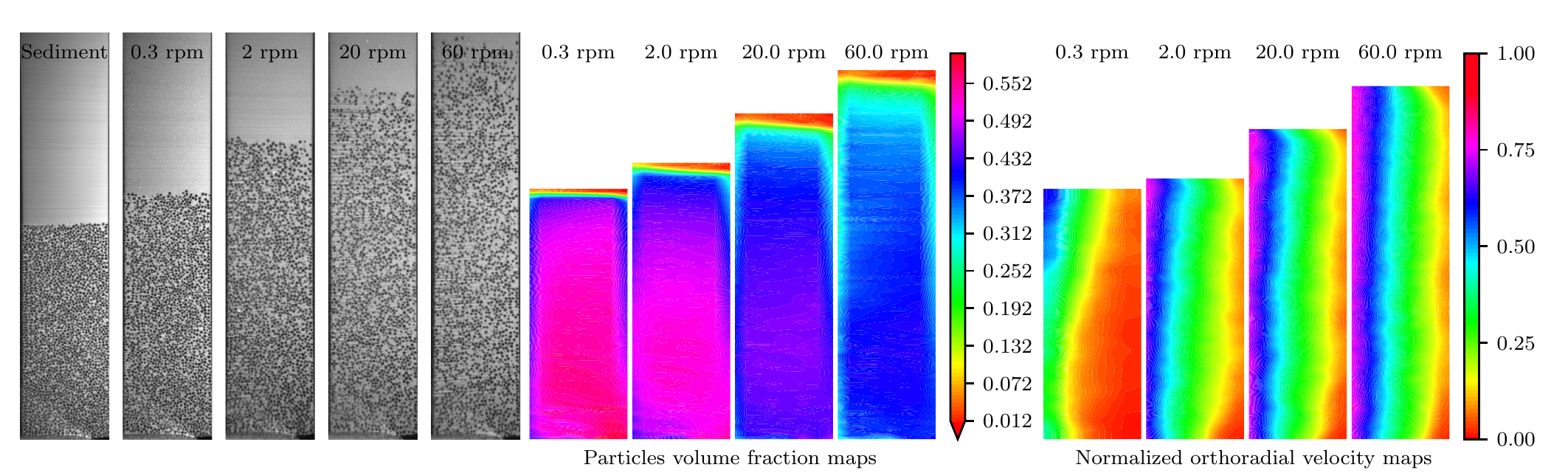}
\caption{Left: typical images recorded for different rotor rotation speeds. A photo of the settled layer ($\Omega=0$) is also presented (height $21.3\, mm$). The rotor is on the left of each frame, the stator is on the right, and the mercury/suspension interface corresponds to the bottom of each frame. Centre: the mapping of the particle volume fraction averaged over 10000 images. Right: the orthoradial velocity normalized by the rotor velocity $\Omega R_1$ and averaged over 100 velocity fields.}
    \label{fig:cartograhie}
\end{figure}

To reach the steady state, the suspension is first sheared with an angular velocity of the rotor equal to $5\,rpm$ for one hour. Then, the speed is set to the desired value for a period until the steady state is reached; the steady stated is considered attained when the torque applied by the rheometer becomes constant. The time duration necessary to achieve the steady state is approximately a few hours.
Figure \ref{fig:cartograhie} shows the viscous resuspension observed for a few rotor angular velocity values. As $\Omega$ increases, the resuspended height increases and the bulk particle concentration decreases. For the slowest rotation speeds (see figure \ref{fig:cartograhie}), particle layering appears near the walls. This structuring of the suspension is clearly observed by averaging images (see figure \ref{fig:glissement}(b)).

\subsection{Velocity and concentration fields}

\subsubsection{Settled layer}
Figure \ref{fig:cartograhie} shows an image of the suspension in the settled state. The sediment height is $h_0=21.3\, mm\approx 4(R_2-R_1)$. The volume fraction of the sediment was measured in a graduated cylinder approximately $1\, cm$ in diameter. A given mass of particles is poured into the vessel containing the suspending liquid (Triton X100), and the sediment height is carefully measured. We took three measurements and obtained $\phi_0=0.574\pm0.003$.

\subsubsection{Concentration fields}
To determine the concentration field in the $(x,z)$ vertical laser plane, each image is binarized with a local threshold. The particles are detected through a watershed segmentation process \citep{vincent1991watersheds}. The position of the barycentre of each segmented zone gives the position of each particle centre in the $(x,z)$ plane sampled with rectangular cells $(i,j)$ of size $\delta x=(R_2-R_1)/8$ and $\delta z=2a$. In each cell, the number of particle centres, $N_{ij}$, is measured. The particle density $n_{ij}=N_{ij}/(\delta x \delta z)$ is reconstructed in the $(r,z)$ plane, making the change of variable $r=\sqrt{y_0^2+x^2}$. Due to the non-zero thickness of the laser sheet and of the slight polydispersity of the particles, $n_{ij}$ is not the absolute particle density, and to compute the true particle volume fraction, we use the particle volume conservation from the sediment to the resuspended state:
\begin{equation}
  \phi(r,z)=\beta n(r,z) \mbox{ with }\beta= \dfrac{\phi_0 \pi ( R_2^2-R_1^2 )h_0}{\int_0^h \int_{R_1}^{R_2} n_{ij} 2\pi r\, \mathrm{d}r \mathrm{d}z }
 \end{equation}

To compute the mean particle volume fraction, this procedure is repeated over $10000$ decorrelated images. Note that the acquisition time can be as long as $100\,hrs$ for the lowest rotation speed of the rotor.
Examples of the concentration field are given in figure \ref{fig:cartograhie}, which deserves a few comments:
\begin{itemize}
\item Near the walls, the particle fraction is lower than in the bulk of the suspension, which should stem from the layering of the particles near the walls \citep{Suzuki2008Study, Yeo2010Ordering, blanc2013microstructure, gallier2016effect, deboeuf2018imaging}.
\item Outside of the structured zones, no or very weak radial particle migration is observed: the maximum difference in the particle volume fraction is evaluated to be less than $2\%$.
\item Along the vertical direction, a concentration gradient is observed as expected in the case of resuspension flows with a sharp interface separating the suspension and the pure fluid \citep{acrivos1994measurement}.
\end{itemize}

\subsubsection{Velocity fields}
The aim of the present study is to investigate resuspension and to link it to particle normal stresses. Because $\Sigma_{33}^p$ is a function of the shear rate, it is essential that the shear rate is known as precisely as possible. For this purpose, we measured the velocity field in the gap. The shift in the laser sheet out of the radial plane allows particle image velocimetry (PIV) measurements \citep{manneville2004high} in the $(x,z)$ plane. Under the assumption that the radial component of the velocity is zero or much smaller than the orthoradial component, $v_{\theta}$ can be deduced from a simple projection of  $v_{x}$ along the orthoradial direction (see figure \ref{fig:schema}(b)):
\begin{equation}
v_\theta(x,z)=v_x(x,z) \dfrac{\sqrt{x^2+y_0^2}}{y_0}
\end{equation}
The velocity field $\boldsymbol{v}(v_x(x,z),v_z(x,z))$ is computed using the open source software DPIVSOFT\footnote{Available on the web (https://www.irphe.fr/meunier/)} \citep{meunier2003analysis}. Each image is divided into correlation windows of size $128 \times 128\, px^2$. Each correlation window contains approximately $10$ particles that are the PIV tracers. The cross correlation of the corresponding windows from two successive images yields the mean velocity of the particles in the window. The in-plane loss of pairs error is decreased by translating the correlation windows in a second run \citep{westerweel1997fundamentals}, thus reducing the correlation windows size to $64 \times 64\, px^2$. The same procedure performed on all the windows gives the velocity field, which is averaged over $100$ images.

The mapping of the $\theta$-component of the velocity field in the plane $(x,z)$ is then obtained and used to reconstruct the velocity field in the $(r,z)$ plane. Velocity maps are shown in figure \ref{fig:cartograhie}, in which the velocity normalized by the velocity of the rotor is represented for several values of $\Omega$. Note that the PIV measurements also enable estimation of the z-component of the  velocity, particularly to check that there is no significant secondary flow. For all the experiments, we checked that the vertical velocity was less than $1\%$ of $v_\theta$ and did not present any peculiar spatial correlation.
It is clear from figure \ref{fig:cartograhie} that there is a significant apparent wall slip, especially for the low angular velocities, i.e., the large particle volume fractions. The wall slip phenomenon in concentrated non-Brownian suspensions is well known \citep{jana1995apparent,ahuja2009slip,blanc2011local,korhonen2015apparent} and can be at the origin of the very large discrepancy between the macroscopic expected shear rate, $\dot{\gamma}_N=2\Omega \dfrac{R_1^2 R_2^2}{R_2^2-R_1^2} \dfrac{1}{r^2}$, and the true local shear rate that can be deduced from the PIV measurements:
\begin{equation}
  \dot{\gamma}(r,z)=r \dfrac{\partial (v_\theta(r,z)/r)}{\partial r}
  \label{gamma_dot}
\end{equation}

\begin{figure}
    \centering
    \includegraphics[width=0.6\textwidth]{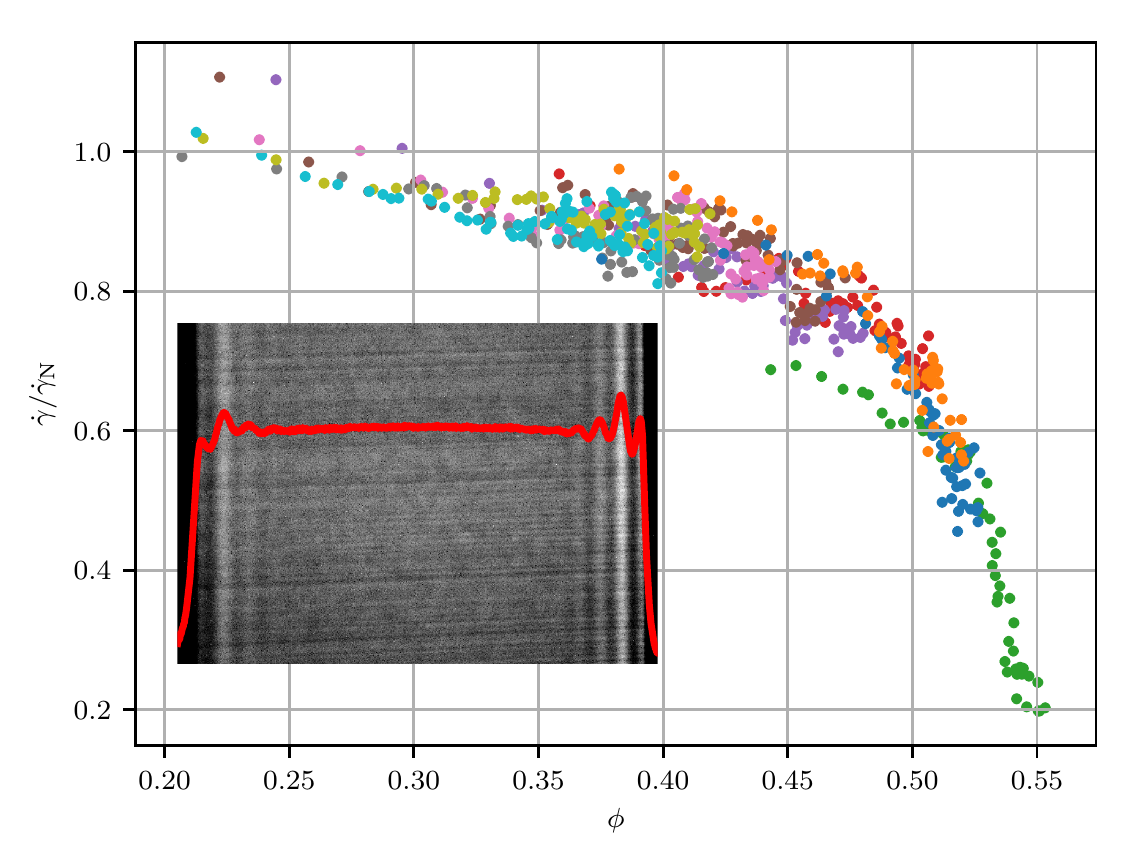}
\caption{Ratio of the local shear rate to the nominal shear rate vs. the local volume fraction. Each colour corresponds to a given value of $\Omega$. Each point was obtained by averaging $\dot{\gamma}$ and $\phi$ over the central third of the gap for a given height $z$. $\dot{\gamma}_N$ is calculated at the middle of the gap $r=(R_1 + R_2)/2$. Insert: a zoomed-in image of an averaged image over 10000 decorrelated frames obtained for $\Omega = 2\, rpm$. Red line: spatial variation in the particle volume fraction (a.u.) }
    \label{fig:glissement}
\end{figure}

Figure \ref{fig:glissement} displays the ratio of the measured shear rate to the nominal shear rate calculated at the middle of the gap, $\dot{\gamma}_N(r = (R_1+R_2)/2)$, as a function of $\phi$ for all the values of $\Omega$. A few comments on this figure are needed. First, it is observed that all the data collapse on a unique curve regardless of the angular velocity of the rotor. This finding is consistent with the results of \cite{jana1995apparent} and the fact that the difference between $\dot{\gamma}$ and $\dot{\gamma}_N$ is mainly due to apparent wall slip that arises from particle layering near the cylinders \citep{blanc2013microstructure}. Particle layering is clearly seen in figure \ref{fig:glissement}(inset), which is an averaged image obtained for $\Omega=2\,rpm$. Second, wall slip becomes negligible when the particle volume fraction is small enough $\phi\approx0.2$. In contrast, for higher concentrations, the local shear rate can substantially deviate from $\dot{\gamma}_N$; for the smallest values of $\Omega$ (the largest values of $\phi$), the true shear rate can be as small as one-fourth of the apparent macroscopic shear rate, making it necessary to measure the velocity field in the gap.

\subsection{Results}
\subsubsection{Concentration profiles}
\begin{figure}
    \centering
    \includegraphics[width=0.6\textwidth]{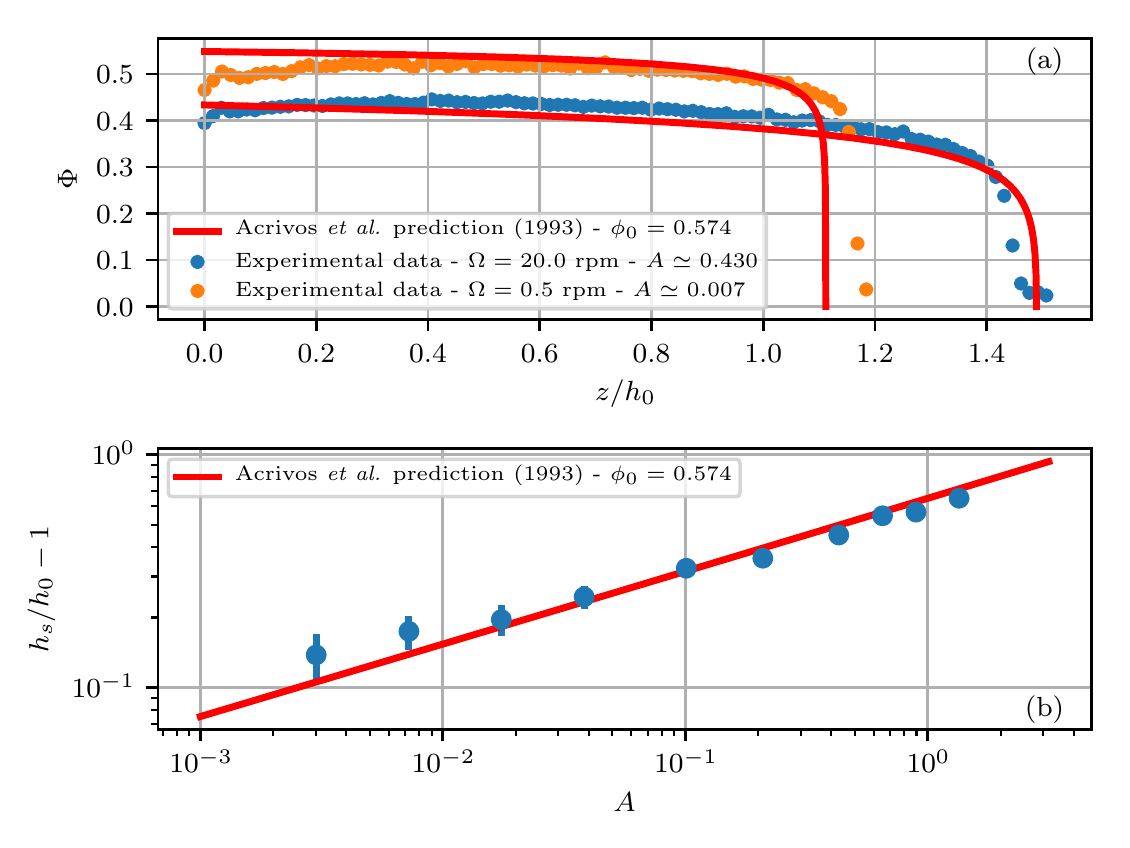}
\caption{a) Examples of the vertical concentration profiles obtained by averaging $\phi(r)$ over the central third of the gap. The corresponding Shields numbers are computed using the local shear rate (averaged over the central third), and the results are compared to the predictions of \cite{acrivos1993shear} (red lines). b) Relative expansion of the particle layer versus the Shields number. $h_s$ is arbitrary defined as the height in which $\phi=0.1$, and the results are compared with the correlation proposed by \cite{acrivos1993shear}.}
    \label{fig:concentration_profile}
\end{figure}

Figure \ref{fig:concentration_profile}(a) shows two concentration profiles averaged over the central third of the gap at low ($\Omega=0.5\, rpm$) and high ($\Omega=20\, rpm$) angular velocities. It is observed that the concentration is almost constant in the resuspended layer and drops to zero quite sharply, even for the highest angular velocity. This sharp interface between the resuspended layer and the clear fluid was already predicted by \cite{acrivos1993shear} when interpreting their experiments in light of a diffusive flux model. Figure \ref{fig:concentration_profile}(a) also shows the profiles predicted by \cite{acrivos1993shear}. The agreement is quite good even though the resuspension height that we measured at low angular velocity is slightly larger than that obtained by \cite{acrivos1993shear} and marginally smaller at high angular velocity. This trend is seen in figure \ref{fig:concentration_profile}, where, as in \cite{acrivos1993shear}, we observe a power-law dependence of the sediment expansion with the Shields number (\ref{A}) but with an exponent slightly lower than $1/3$ \citep{zarraga2000characterization}.

Finally, it should be noted that near the bottom of the Couette cell, the particle concentration tends to decrease. This finding may be related to a problem of particle detection near the interface with mercury, which reflects light and may downgrade the image quality in its vicinity. In the next section, in which the concentration will be used to evaluate $\Sigma_{33}^p$, we will not consider this zone in which we are not absolutely confident in the particle concentration measurement.

\subsubsection{Determination of $\Sigma_{33}^p$}
To determine $\Sigma_{33}^p$, Eq.\ref{Cauchy} is integrated from the interface between the resuspended layer and the clear fluid to the height $z(\phi)$:

\begin{equation}
    \Sigma_{33}^p(r,z)=-\int_{z(\phi)}^{z(\phi=0)}\Delta \rho g \phi(r,z) dz
\end{equation}

\begin{figure}
    \centering
    \includegraphics[width=0.6\textwidth]{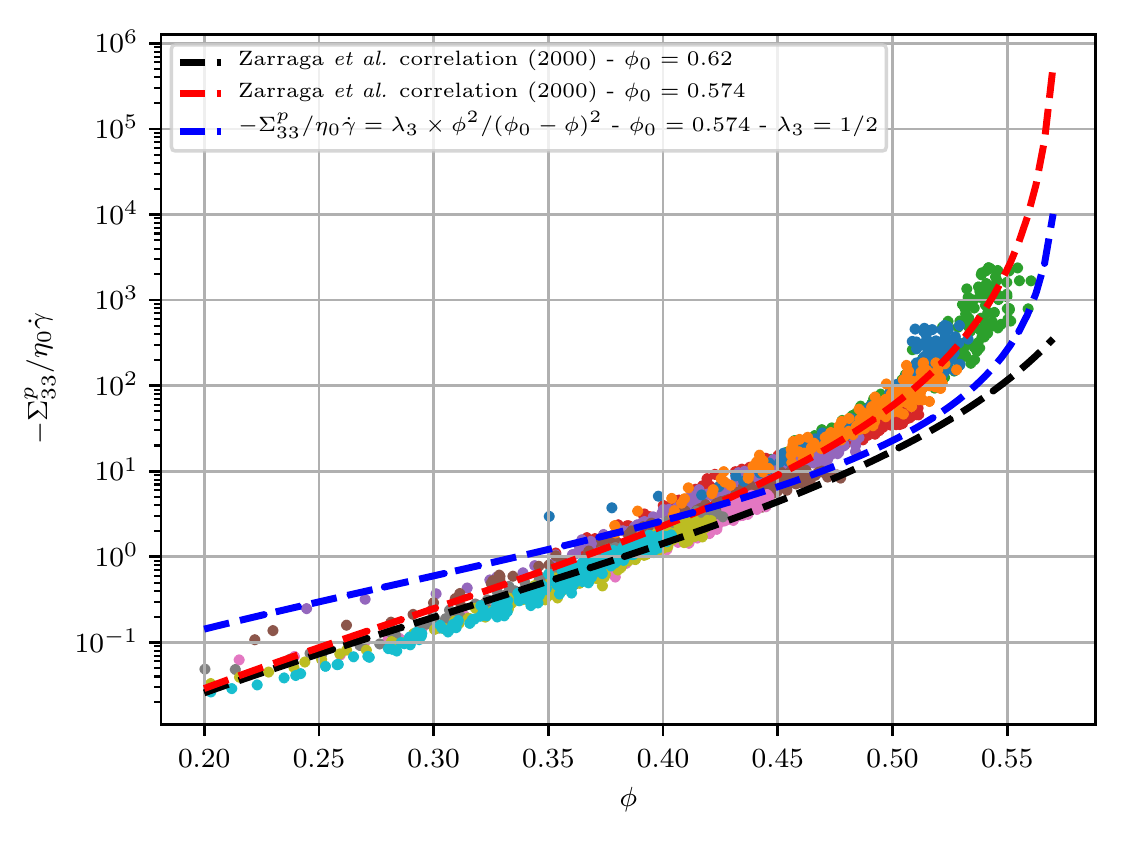}
\caption{Third particle normal stress normalized by the product of the fluid viscosity and the local shear rate versus the local particle volume fraction. The agreement with the correlation proposed by \cite{zarraga2000characterization} where $\phi_0=0.574$ is very good (red line). In contrast, it is not possible to represent the experimental results by the correlation obtained by \cite{boyer2011unifying} for $\Sigma_{22}^p$ together with $\lambda_3=1/2$ (blue line).}
    \label{fig:sigma33}
\end{figure}

Thus, for each point $(r,z)$, the third particle normal stress, the local shear rate and the particle volume fraction are computed.
Figure \ref{fig:sigma33} shows the variation of $-\Sigma_{33}^p$ normalized by $\eta_0 \dot{\gamma}$ as a function of $\phi$. To avoid boundary effects, we discarded the measurements taken for $z<h_s/4$ and $r<(R_2-R_1)/3$ or $r>2(R_2-R_1)/3$.

In figure \ref{fig:sigma33}, we observe that the data almost collapse on a single curve for a wide range of $\phi$ between $0.2$ and $0.6$ with variation in $\Sigma_{33}^p/\eta_0 \dot{\gamma}$ over more than five decades. We restricted the data to particle volume fractions greater than $0.2$ because, as shown in figure \ref{fig:concentration_profile}, below this value, the concentration profile is very sharp, which makes it difficult to measure the concentration. The data are somewhat scattered, especially for the largest values of $\phi$. This finding may have different origins. First, it can stem from experimental issues because, as observed in figure \ref{fig:concentration_profile}, for the lowest angular velocity values (i.e., the larger particle volume fractions), the profile is nearly flat, which means that $\Sigma_{33}^p$ given by the integral of $\phi(z)$ varies greatly with $\phi$. Thus, even a small error in $\phi$ is likely to cause a large error in the computation of $\Sigma_{33}^p$. In addition to these experimental issues, the shear-thinning character displayed by most non-Brownian suspensions \citep{acrivos1994measurement,lobry2019shear,vazquez2016shear,vazquez2017investigating,tanner2018bootstrap} can also be at the origin of the scattering observed for the largest concentrations. Indeed, it is expected that particle normal stresses 
vary with the shear stress (and not the shear rate) \citep{boyer2011unifying}. Thus, as the angular velocity increases, the viscosity, $\eta_S$, decreases and the ratio $\Sigma_{33}^p/\eta_S$ increases for a given $\phi$, which should improve the collapse of the data.

The red and black lines in figure \ref{fig:sigma33} represent the correlation proposed by \cite{zarraga2000characterization}:

\begin{equation}
    \Sigma_{33}^p=-\eta_0 \dot{\gamma} \dfrac{\phi^3}{\left(1-\dfrac{\phi}{\phi_0}\right)^3}
\end{equation}

The black curve is obtained with the original value of $\phi_0$ proposed by \cite{zarraga2000characterization} ($\phi_0=0.62$), while the red curve has been obtained for $\phi_0=0.574$: the value of the particle volume fraction inside the settled layer that we measured. We observe a very good agreement between the
experimental data and the correlation from \cite{zarraga2000characterization}. Furthermore, \cite{zarraga2000characterization} established the correlation for a particle volume fraction ranging from $0.3$ to $0.5$, while our results show that this correlation can be expanded to a wider range of concentrations.
The blue curve is obtained by using the correlation obtained by \cite{boyer2011unifying} for $\Sigma_{22}^p$ (with $\phi_0=0.574$):
\begin{equation}
    \Sigma_{22}^p=-\eta_0 \dot{\gamma}\dfrac{\phi^2}{\left(\phi_0-\phi\right)^2}
\end{equation}
and assuming that
$\lambda_2=\Sigma_{22}^p/\Sigma_{11}^p\approx 1$ and  $\lambda_3=\Sigma_{33}^p/\Sigma_{11}^p=0.5$, as suggested by \cite{morris1999curvilinear}.

The agreement between the blue curve and our data is not satisfactory. In our opinion, this discrepancy does not call into question the results obtained by \cite{boyer2011unifying} but rather the lack of variability in $\lambda_{3}$ with $\phi$. This last result has already been noted by \cite{gallier2014rheology} and was previously suggested by \cite{morris1999curvilinear} themselves.

In conclusion, with the aim of studying viscous resuspension, we conducted local measurements of both the shear rate and the particle volume fraction and deduced the variation of $\Sigma_{33}^p/\eta_0\dot{\gamma}$ with $\phi$. Our results confirm the correlation proposed by \cite{zarraga2000characterization} and extend it over a wider range of particle volume fractions.

We are grateful to L. Lobry, F. Peters and B. Saint-Michel for fruitful discussions, and D. Gilbert for the 3D sketch of the experimental device.

\bibliographystyle{jfm}

\bibliography{biblio_resuspension.bib}

\end{document}